\documentclass[aps,prl,preprintnumbers,groupedaddress,twocolumn]{revtex4-2}
\usepackage{amsmath}
\usepackage{booktabs}
\usepackage{hyperref}
\usepackage{graphicx}
\usepackage{siunitx}

\begin{document}

\preprint{CERN-TH-2025-189}
\preprint{MITP-25-060}

\title{Non-singlet vector current in lattice QCD: $\Oa$-improvement from large volumes}

\author{Tim~Harris}
\affiliation{Institute for Theoretical Physics, ETH Z\"urich,\\Wolfgang-Pauli-Strasse 27, 8093 Z\"urich, Switzerland.}

\author{Harvey~B.~Meyer}
\affiliation{Theoretical Physics Department, CERN, 1211 Geneva 23, Switzerland}
\affiliation{PRISMA+ Cluster of Excellence \& Institut f\"ur Kernphysik, Johannes Gutenberg-Universit\"at Mainz, D-55099 Mainz, Germany}
\affiliation{Helmholtz Institut Mainz, Staudingerweg 18, D-55128 Mainz, Germany}

\newcommand{\mbar}{\bar m}
\newcommand{\Oa}{\mathrm O(a)}
\newcommand{\Nf}{N_\mathrm{f}}
\newcommand{\cv}{c_\mathrm{V}}
\newcommand{\cvtilde}{c_{\tilde{\mathrm{V}}}}
\newcommand{\ba}{b_\mathrm{A}}
\newcommand{\bg}{b_\mathrm{g}}
\newcommand{\bX}{b_X}
\newcommand{\bXbar}{\bar b_X}
\newcommand{\bXtilde}{\tilde b_X}
\newcommand{\Zx}{Z_X}
\newcommand{\Zv}{Z_\mathrm{V}}
\newcommand{\ZA}{Z_\mathrm{A}}
\newcommand{\ca}{c_\mathrm{A}}
\newcommand{\babar}{\bar{b}_{\mathrm{A}}}
\newcommand{\batilde}{\bar {b}^\mathrm{eff}_{\mathrm{A}}}

\date{\today}

\begin{abstract}
    In previous work, we determined the improvement coefficients $\cv$ and
    $\cvtilde$ required for the massless $\Oa$-improvement of the local and
    point-split discretizations of the non-singlet vector current for $\Nf=3$
    non-perturbatively $\Oa$-improved Wilson fermions and the L\"uscher-Weisz
    gauge action, using ensembles of large-volume configurations generated by
    the Coordinated Lattice Simulations (CLS) initiative.
    A new estimate for the mass-dependent improvement coefficient $\batilde$
    has recently become available, differing from the one used in our earlier
    study, and on which our implementation via a massive axial Ward identity
    relied.
    Here, we update our analysis of the mass-independent vector improvement
    coefficients based on the new axial current improvement coefficient, and
    analyse additional ensembles with a different chiral trajectory in order to
    validate our results at two values of the bare coupling.
    We find that using the new estimate of $\batilde$ improves the consistency
    between the two chiral trajectories, as well as with a previous
    determination of the improvement coefficients directly in the massless
    limit on small volumes.
\end{abstract}

\maketitle

\newpage

\section{Introduction}
\label{sec:intro}

Numerical simulation of lattice field theory is one of the most successful and
universal frameworks for studying strongly-coupled quantum field theories from
first principles.
Lattice quantum chromodynamics (QCD) simulations have reached a maturity where
they are essential for the precision study of the Standard Model of Particle
Physics (SM) and in addition provide new insights into exotic states of
matter.
Such computations are often said to be systematically improvable, a feature
which can be observed in the progress of precision lattice predictions over
time, for example in those compiled by the Flavour Lattice Averaging
Group~\cite{FlavourLatticeAveragingGroupFLAG:2024oxs}.

One sense in which lattice simulations are systematically improvable is
realized by Symanzik's improvement
program~\cite{Symanzik:1983gh,Symanzik:1983dc}.
According to effective field theory analysis, the approach to the continuum
limit of renormalized on-shell matrix elements of local operators can be
accelerated by adding higher-dimension counterterms which cancel the
discretization effects order by order in the lattice spacing $a$, so long as
their coefficients are chosen appropriately.
In this way, the uncertainty due to extrapolation to the continuum can be
reduced in a fairly predictable way.

In Wilson's formulation of lattice QCD~\cite{Wilson:1974sk,Wilson:1975id}, due
to the loss of chiral symmetry, the Symanzik analysis predicts leading linear
effects for local operators, at least classically.
In this case, the implementation of the $\Oa$ improvement programme is crucial
to reach the required precision to test the SM with the currently achievable
range of lattice spacings.
The non-perturbative determination of improvement coefficients is mandatory in
the range of couplings used for large-volume simulations, and chiral Ward
identities provide natural and practical conditions for restoring chiral
symmetry and removing the linear lattice effects~\cite{Luscher:1996ug}.
In practice, the improvement of the Wilson fermion action is known for
$\Nf=2,3$ and $4$ flavours and various gauge
actions~\cite{Jansen:1998mx,JLQCD:2004vmw,Cundy:2009yy,Bulava:2013cta,Fritzsch:2018kjg},
but in addition to improving the action, each local operator must, in general,
be improved by adding the necessary counterterms and coefficients, classified in
Ref.~\cite{Bhattacharya:2005rb} for fermionic bilinears.
For $\Nf=3$ $\Oa$-improved Wilson fermions, mass-independent improvement
coefficients have been determined non-perturbatively for many such
operators~\cite{Bulava:2015bxa,Gerardin:2018kpy,Heitger:2020zaq,Chimirri:2023ovl,JonesPetrak:2025kjm},
including the non-singlet axial and vector currents.
At finite quark masses, the number of improvement coefficients proliferates,
the so-called $b$-terms, although some combinations are known
non-perturbatively in the $\Nf=3$
case~\cite{Korcyl:2016ugy,Fritzsch:2018zym,Gerardin:2018kpy,deDivitiis:2019xla,DallaBrida:2023fpl}.
In general, the determination of the improvement coefficients poses a
challenge to obtaining physically-interesting quantities in the continuum.

Correlation functions involving the quark electromagnetic current are of utmost
importance to understand the interactions and structure of hadrons,
fundamental quantities like the hadronic vacuum polarization and
even isospin-breaking corrections.
In particular, then, the $\Oa$ improvement of the vector current is extremely
relevant input for current precision physics programmes.
Following the proposal of Guagnelli and Sommer~\cite{Guagnelli:1997db}, there
are two existing determinations of the mass-independent improvement
coefficients $\cv$ and $\cvtilde$ for the local and point-split
discretizations for $\Nf=3$ $\Oa$-improved Wilson
fermions~\cite{Gerardin:2018kpy,Heitger:2020zaq}.  Although these
determinations display a tension, they are compatible with higher-order
ambiguities.
In this work, we update the former analysis of the mass-independent
improvement coefficients in light of new results for the mass-dependent
improvement on which the axial Ward identity relies.
This improves the consistency between the two determinations dramatically, and
here we provide another determination at two values of the bare coupling using
a different chiral trajectory, adding further confidence to the new
determination.

\section{Background}
\label{sec:bg}

For a theory with more than two quark flavours, a condition from which the
improvement of the vector current can be obtained is the (continuum) chiral
Ward identity~\cite{Guagnelli:1997db} 
\begin{align}
    \int\mathrm d^3 y \,\langle
    \delta S A^{23}_{k}(y) O^{31}(0) \rangle
    &= \int\mathrm d^3y \,\langle
    V_{k}^{13}(y) O^{31}(0) \rangle,
    \label{eq:chiral}
\end{align}
where the variation of the action under the flavoured local axial
transformations $\delta\psi^1(x)=-\alpha(x)\gamma_5\psi^2(x)$ and
$\delta\bar\psi^2(x) = -\bar\psi^1(x)\alpha(x)\gamma_5$ is given by
\begin{align}
    \delta S &= \int_{t_1}^{t_2}\mathrm dx_0\,\int\mathrm d^3x\,
    \{\partial_\mu A_\mathrm{\mu}^{12}(x) - 2m^{12} P^{12}(x) \},
    \label{eq:delaction}
\end{align}
and $m^{rs}=\tfrac{1}{2}(m^r+m^s)$.
In writing eq.~\eqref{eq:delaction} we have chosen the transformation
$\alpha(x)$ to be unity in the domain $t_1<x_0<t_2$ and null everywhere else.
As the preceding Ward identity is violated in Wilson's lattice formulation by
$\mathrm O(a)$ cutoff effects, it can be used as a condition to fix one of
the improvement coefficients which contribute, provided that the remaining
coefficients are set appropriately to cancel the other effects at that order.

To this end, we must specify the discretization of the operators and
integral in the chiral Ward identity.
We write the massless $\mathrm O(a)$-improved axial and vector currents for
$r\neq s$
\begin{align}
    A^{rs}_{\mathrm{I},\mu}(x) &= A_\mu^{rs}(x)
    + a\ca\tilde\partial_\mu P^{rs}(x),\\
    V^{rs}_{\mathrm{I},\mu}(x) &= V_\mu^{rs}(x)
    + a\cv\tilde\partial_\nu T^{rs}_{\mu\nu}(x),
    \label{eq:impcurrents}
\end{align}
in terms of the bare currents
\begin{align}
    A_\mu^{rs} &= \bar\psi^r\gamma_\mu\gamma_5\psi^s,\qquad
    V_\mu^{rs} = \bar\psi^r\gamma_\mu\psi^s,
    \label{eq:barecurrents}
\end{align}
and the pseudoscalar density and tensor operator
\begin{align}
    P^{rs} &= \bar\psi^r\gamma_5\psi^s,\qquad
    T_{\mu\nu}^{rs} = \bar\psi^r\sigma_{\mu\nu}\psi^s
    \label{eq:densities}
\end{align}
where the antisymmetric tensor matrix is defined as
$\sigma_{\mu\nu}=-\tfrac{1}{2}[\gamma_\mu,\gamma_\nu]$.
The finite difference operator $\tilde\partial_\mu$ denotes the symmetric $\mathrm O(a)$-improved definition.

In addition, we also compute the massless improvement for the point-split
vector current
\begin{align}
    \nonumber
    \hat V^{rs}_\mu(x) &= \tfrac{1}{2}\{\bar\psi^r(x+a\hat\mu)U_\mu^\dagger(x)
        (1+\gamma_\mu)\psi^s(x)\\
    &- \bar\psi^r(x)(1-\gamma_\mu)U_\mu(x)\psi^s(x+a\hat\mu)\},
    \label{eq:pscurr}
\end{align}
or more precisely the site-centered version $\tilde
V^{rs}_\mu(x)=\tfrac{1}{2}(\hat V^{rs}_\mu(x)+\hat V_\mu(x-a\hat\mu))$, that
has the same lattice symmetries as the local current, which allows us to use
the same form for the improvement counterterm as for the local current in
eq.~\eqref{eq:impcurrents}, but with coefficient
$\cvtilde$~\cite{Frezzotti:2001ea}.

In the massive theory, there are additional mass-dependent counterterms which
are proportional to the bare operator and therefore can be conveniently
parameterized together with any multiplicative renormalization $Z_{J}$ for
massless-improved bilinears $J^{rs}_\mathrm{I}$ with $r\neq s$
as~\cite{Bhattacharya:2005rb}
\begin{align}
    J_\mathrm{R}^{rs} = \hat Z_{J}
    J_\mathrm{I}^{rs} = Z_{J}(\tilde g_0^2)
    (1+\bar b_{J} \Nf a\bar m + b_{J} am^{rs})
    J_\mathrm{I}^{rs},
    \label{eq:mdepimp}
\end{align}
where $\bar m$ is the average bare subtracted quark mass.
In the massive theory, there is one additional sticking point, which is that
the bare coupling also needs to be improved
\begin{align}
    \tilde g_0^2 = g_0^2(1+b_\mathrm{g} a\bar m),
    \label{eq:impcoupling}
\end{align}
in order to fix the lattice spacing for different quark masses.
Given that a shift in the bare coupling of the coefficients multiplying
$\mathrm O(a)$-improvement terms will only contribute at higher orders in $a$,
we only need to care about the dependence of the renormalization factor on
this coupling.
In fact, for the currents it is easy to see that defining
$\bar b_{J}^\mathrm{eff}=\bar b_{J} + b_\mathrm{g}/\Nf
\{\mathrm d\ln Z_{J}/\mathrm d\ln g_0^2\}$, we can absorb the
dependence on $b_\mathrm{g}$ into $\bar b^\mathrm{eff}_{J}$ %
\footnote{A non-perturbative determination of $\bg$ has recently become
available although not directly in the regime of couplings relevant for the
CLS simulations~\cite{DallaBrida:2023fpl}} so that
\begin{align}
    \hat Z_{J}
    J_\mathrm{I}^{rs} = Z_{J}(g_0^2)
    (1+\bar b_{J}^\mathrm{eff} a\bar m + b_{J} am^{rs})
    \label{eq:effimp}
\end{align}
up to neglected $\mathrm O(a^2)$ effects.
Note that as the point-split current satisfies a vector Ward identity at
finite lattice spacing, the entire normalization is constrained to be unity,
i.e.~$\hat Z_{\tilde{\mathrm V}}=1$.

Inserting these definitions of the lattice operators in the chiral Ward
identity, we arrive at a family of definitions for the improvement
coefficients
\begin{align}
    c_{J} &= \frac{a^3\sum_{\vec y}\langle
    \delta S A_{\mathrm R,k}^{23}(y) O^{31}(0) - \hat Z_{J} J_k^{13}(y) O^{31}(0) \rangle}
    {a^4\sum_{\vec y} \langle \hat Z_{J}\tilde\partial_\nu T^{13}(y)O^{31}(0)\rangle}
    \label{eq:cvimpcond}
\end{align}
for $J={V,\tilde V}$ and with
\begin{align}
    \nonumber
    \delta S &= a^3\sum_{\vec x}(A^{12}_{\mathrm R,0}(t_2,\vec
    x)-A^{12}_{\mathrm R,0}(t_1,\vec x)) \\
    &\qquad\qquad - a^4\sum_{x_0=t_1}^{t_2}\sum_{\vec x} 2m^{12}P^{12}(x),
    \label{eq:latact}
\end{align}
which depend on $t_1$, $t_2$, $y_0$, the choice of $O^{31}$ and $m^j$.
We note that, although the last term is renormalized and scale-independent, a
contact term arises when it coincides with $y$ which must be removed by
sending $m^{12}$ to zero.
Therefore, with $\Nf=3$, different choices for $m^3$ in this limit, or
correspondingly $\bar m$, give rise to different valid definitions of the
improvement coefficients.
A non-zero value of $m^3$, however, requires that the axial current is
improved in the massive theory and therefore the input of the correct
mass-dependent improvement coefficients.
While in earlier work we used an improvement condition with non-zero $\bar m$
in the limit of $m^{12}\rightarrow0$, in this work we add a determination with
$\bar m=0$ for two of the values of the lattice spacing.
This provides a strong cross-check on the validity of the implementation in
the massive case.

\section{Numerical determination}

The definition of the improvement condition presented in the previous section
is the same as in Ref.~\cite{Gerardin:2018kpy}.
In addition, we choose the same values of the coordinates $t_1$, $t_2$ and
$y_0$ as used there.
To ensure a smooth approach to the continuum limit, we fix the kinematic
variables in the improvement condition in physical units to the same values
used in the earlier work with $t_2-t_1\approx0.7\,\mathrm{fm}$ while the
current was placed at distance $y_0\approx0.8\,\mathrm{fm}$ from the external
operator.
The external operator is also unchanged, and in particular, we use the sum of
the local vector current and tensor operator as in that work.
Firstly as in our previous work, we impose the improvement coefficient at vanishing
$m^{12}$ and with $\bar m=\bar m^{\mathrm{phys}}$ and secondly we add a
determination with $\bar m=0$, by extrapolating from the ensembles with
$m^1=m^2=m^2$ listed in Tab.~\ref{tab:ens}.
The massive vector renormalization factor is the same as used in
Ref.~\cite{Gerardin:2018kpy} determined by imposing a vector Ward
identity.

\begin{table}[t]
    \centering
    \begin{tabular}{ccccc}
        \toprule
        $6/g_0^2$ & id & $\kappa$ & $L_0/a\times (L/a)^3$ & $m_\pi$ (MeV) \\
        \midrule
        3.46 & rqcd29 & 0.136600  & $64\times 32^3$ & 710 \\
             & rqcd30 & 0.1369587 & $64\times 32^3$ & 320 \\
             & X450   & 0.136994  & $64\times 48^3$ & 260 \\
        \midrule
        3.55 & X250 & 0.137050 & $64\times 48^3$ & 350 \\
             & X251 & 0.137100 & $64\times 48^3$ & 270 \\
        \bottomrule
    \end{tabular}
    \caption{New ensembles analysed in this work with $m^{12}=m^3$, in
        addition to those presented in Ref.~\cite{Gerardin:2018kpy} made
        available through the CLS initiative~\cite{Bruno:2014jqa}.
     The ensembles labelled by ``rqcd'' were generated by the Regensburg
 group~\cite{Bali:2023sdi} with the BQCD code~\cite{Nakamura:2010qh}.}
    \label{tab:ens}
\end{table}

\subsection{Analysis using updated \texorpdfstring{$\batilde$}{bAtilde}}
\label{sec:newbatilde}

\begin{figure}[t]
    \centering
    \includegraphics[scale=0.65]{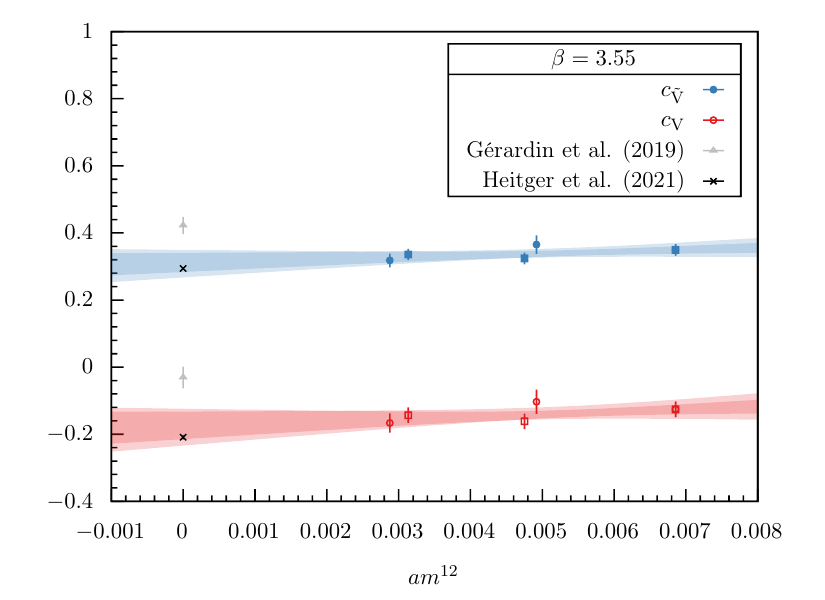}
    \caption{Dependence on the estimator as a function of $m^{12}$ at fixed
        bare coupling $\beta=6/g_0^2=3.55$ for the local (red) and point-split
        (blue) discretizations of the vector current.
        The squares display the chiral trajectory with $\mbar=0$ while the
        circles show the same with $\mbar=\mathrm{const.}$. 
    The earlier results from Ref.~\cite{Gerardin:2018kpy} are shown in the
    grey triangles while the results obtained from SF~\cite{Heitger:2020zaq} are
    shown with crosses.}
    \label{fig:finmass}
\end{figure}

Compared with the previous analysis, we use the updated values for $\batilde$
given in Tab.~\ref{tab:axial} from Ref.~\cite{Bali:2023sdi,Bali:inprep} based
on the methodology of Ref.~\cite{Korcyl:2016ugy}.
The values differ significantly from the set used in the previous work which
ranged between $\batilde=0.4\text{--}2.0$ from the smallest to largest bare coupling.
Even though a generous uncertainty was attributed to their value, it did not
cover the range of the new determinations, which are now compatible with zero.
Using these new values, we observe a significant decrease of the estimates for
both improvement coefficients at finite mass compared with the previous
determination.
The $m^{12}\rightarrow0$ limit is very flat, and we can obtain reasonable
parameterizations either with a constant or linear behaviour, see
Fig.~\ref{fig:finmass}.
The determination with $\bar m=0$ (obtained from the red squares) is
independent of the values of the mass-dependent improvement coefficients, and
we checked that even if they are not set to their correct values, the limit is
consistent, albeit reached with a larger slope.
The resulting values from the trajectory with $\bar m=\mathrm{const.}$ (blue
circles) are in good agreement with the former.
Both determinations with $m^{12}=0$ are listed in Tab.~\ref{tab:cv_beta}.

\begin{table}[t]
    \centering
    \begin{tabular}{ccccc}
        \toprule
        $6/g_0^2$        & 3.4       & 3.46     & 3.55      & 3.7        \\
        \midrule
        $\ba     $       & 1.244     & 1.239    & 1.232     & 1.221      \\
        $\batilde$       & -0.11(13) & 0.08(11) & -0.03(13) & -0.047(75) \\
        \bottomrule
    \end{tabular}
    \caption{Values of the axial improvement coefficients used in this
    work~\cite{Bali:2023sdi,Bali:inprep}.}
    \label{tab:axial}
\end{table}

\begin{table}[t]
    \centering
    \begin{tabular}{ccccc}
    \toprule    
                & \multicolumn{2}{c}{$\cv$} & \multicolumn{2}{c}{$\cvtilde$} \\
    $6/g_0^2$   & linear       & constant      & linear      & constant     \\
	\midrule
    3.4         & $-0.252(34)$ & $-0.1888(97)$ & $0.256(24)$ & $0.3029(72)$ \\
	\midrule
    3.46        & $-0.165(39)$ & $-0.148(11)$  & $0.322(27)$ & $0.3334(78)$ \\
    3.46$^\star$ & $-0.159(19)$ & $-0.136(11)$  & $0.324(14)$ & $0.3412(82)$ \\
	\midrule
    3.55        & $-0.183(51)$ & $-0.134(16)$  & $0.304(37)$ & $0.342(12)$  \\
    3.55$^\star$ & $-0.169(45)$ & $-0.143(13)$  & $0.317(33)$ & $0.337(10)$  \\
	\midrule
    3.7         & $-0.107(37)$ & $-0.119(11)$  & $0.367(26)$ & $0.3569(87)$ \\
    \bottomrule
    \end{tabular}
    \caption{Values of the massless $\Oa$-improvement coefficients $\cv$ and
    $\cvtilde$ determined from linear and constant fits to the massive axial
    Ward identity. Both models provide good descriptions of the data.
    The values marked with the $\star$ indicate the chiral limit with
    $\bar m=0$.}
    \label{tab:cv_beta}
\end{table}

\begin{figure}[t]
    \centering
    \includegraphics[scale=0.65]{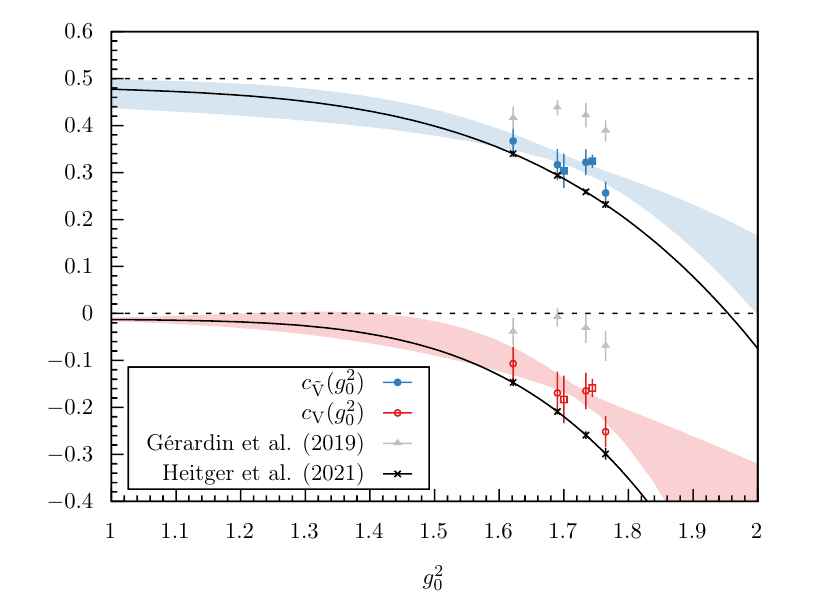}
    \caption{Final results for $\cv$ (red) and $\cvtilde$ (blue) as a function
        of the bare coupling, compared with the results of Heitger et al.
        \cite{Heitger:2020zaq} (black line).
        The old results of Ref.~\cite{Gerardin:2018kpy} are shown with the
    grey triangles.}
    \label{fig:cv_beta}
\end{figure}

\section{Results and discussion}
\label{sec:results}

The results obtained in the previous section are shown as a function of the
bare coupling (blue and red points) in Fig.~\ref{fig:cv_beta}.
They lie systematically between the ones obtained in our earlier
work (grey triangles)~\cite{Gerardin:2018kpy} and
those obtained in small volumes with Schr\"odinger functional (SF) boundary
conditions (black squares)~\cite{Heitger:2020zaq}.
In order to compare systematically with the latter, and because we observe a
sharp decrease of the results over the range of bare couplings, we opt to use
the same parameterization of the data as in that work, which we find to
provide a very good description of the data displayed with the bands.
In addition, the interpolation may be used within or close to the measured
range of values and provides a smooth parameterization of the improvement
coefficient towards the continuum limit.

The parameterization used for the local current is
\begin{align}
    \nonumber
    \cv(g_0^2) &= -0.0130\,C_\mathrm{F} g_0^2\times\\
    &\qquad\qquad
    \Big[1+(a + b g_0^2)\exp\{-(2b_0g_0^2)^{-1}\}\Big]
    \label{eq:interp_loc}
\end{align}
with $C_\mathrm{F}=4/3$ and for the point-split current
\begin{align}
    \cvtilde(g_0^2) &= \tfrac{1}{2} - c g_0^2
\Big[1+ d g_0^2\exp\{-(2b_0g_0^2)^{-1}\}\Big]
    \label{eq:interp_cons}
\end{align}
and the resulting best fit parameters obtained are
\begin{align}
    a &=  -2514.80,\quad b = 2044.53,
    \label{eq:best_fit_loc}
\end{align}
for the local current; respectively, for the point-split one,
\begin{align}
    c &=  316.96,\quad     d = 0.024543.
    \label{eq:best_fit_cons}
\end{align}
The individual parameters are not significantly different from zero, so
including the covariance matrix as given in Tab.~\ref{tab:cov} is essential
to providing a useful and accurate representation of the uncertainty.
Typically, however, the improvement coefficient may be considered part of the
definition of the lattice theory, and its statistical uncertainty may be
considered part of its definition so long as they are determined correctly up to the expected $\Oa$
ambiguity.
In our case, we checked that the difference between our determination and that
of the SF is consistent with such an ambiguity, albeit with a large
uncertainty.

\begin{table}[t]
    \sisetup{
        table-format = 1.2e1,
        round-mode=places,
        round-precision=2,
        table-column-width=6em,
    }
    \centering
    \begin{tabular}{cSSSS}
        \toprule
        $\mathrm{Cov}$ & \multicolumn{1}{c}{$a$} &
        \multicolumn{1}{c}{$b$} & \multicolumn{1}{c}{$c$} & \multicolumn{1}{c}{$d$} \\
        \midrule
        $a$ & 2.14547671e+07   & -1.24111356e+07  & -1.07807454e+10 &  4.78993917e+01\\
        $b$ & -1.24111356e+07  & 7.18200133e+06   & 6.23081218e+09  &  -2.76555769e+01\\
        $c$ & -1.07807454e+10  & 6.23081218e+09   & 4.80661265e+13  &  -9.44972138e+04\\
        $d$ & 4.78993917e+01   & -2.76555769e+01  & -9.44972138e+04 & 9.43765295e-04 \\
        \bottomrule
    \end{tabular}
    \caption{Covariance matrix of best fit parameters for the
    parameterizations for both improvement coefficients for the two
    discretizations of the vector current.}
    \label{tab:cov}
\end{table}

We note that the difference between the improvement coefficients of the two
vector-current discretizations is to a very
good degree consistent with the leading order difference in $g_0^2$, namely
0.5 over the whole range of couplings.
The updated values of these improvement coefficients have already been put to
use in Ref.~\cite{Djukanovic:2024cmq}, where good consistency was observed
between the continuum limit of the integrated current correlator using either
this determination of improvement coefficients or the ones of Heitger and
Joswig~\cite{Heitger:2020zaq}.

\begin{acknowledgements}
We thank Gunnar Bali and Piotr Korcyl for a useful exchange on the improvement
coefficients of the axial current, and Antoine G\'erardin for helpful
discussions.
We also thank Simon Kuberski for providing comments on the first draft
of this manuscript.
We are grateful for the access to the ensembles used here, made available to
us through CLS and our colleagues in the Regensburg group.
Correlation functions were computed on the platforms “Mogon II” at Johannes
Gutenberg University Mainz.
\end{acknowledgements}

\bibliography{cv_update}

\end{document}